\documentclass[aps,prd,twocolumn,superscriptaddress,amsmath,showpacs]{revtex4-1}
\usepackage{graphicx}
\usepackage{dcolumn}
\usepackage{bm}
\usepackage{natbib}

\newcommand{\be}{\begin{equation}}
\newcommand{\ee}{\end{equation}}
\newcommand{\bea}{\begin{eqnarray}}
\newcommand{\eea}{\end{eqnarray}}

\newcommand{\Mpc}{{\rm ~Mpc}}

\begin{document}
\title{Tickling the CMB damping tail: scrutinizing the tension between the ACT and SPT experiments.}

\author {Eleonora Di Valentino}
\affiliation{Physics Department and INFN, Universit\`a di Roma ``La Sapienza'', Ple Aldo Moro 2, 00185, Rome, Italy}

\author {Silvia Galli}
\affiliation{Institut d`Astrophysique de Paris, UMR 7095-CNRS Paris,\\
 Universit\'e Pierre et Marie Curie, boulevard Arago 98bis, 75014, Paris, France}

\author {Massimiliano Lattanzi}
\affiliation{Dipartimento di Fisica e Science della Terra, Universit\`a di Ferrara and INFN, sezione di Ferrara, 
Polo Scientifico e Tecnologico - Edificio C Via Saragat, 1, I-44122 Ferrara Italy
}

\author {Alessandro Melchiorri}
\affiliation{Physics Department and INFN, Universit\`a di Roma ``La Sapienza'', Ple Aldo Moro 2, 00185, Rome, Italy}

\author {Paolo Natoli}
\affiliation{Dipartimento di Fisica e Science della Terra, Universit\`a di Ferrara and INFN, sezione di Ferrara, 
Polo Scientifico e Tecnologico - Edificio C Via Saragat, 1, I-44122 Ferrara Italy
}

\author {Luca Pagano}
\affiliation{Physics Department and INFN, Universit\`a di Roma ``La Sapienza'', Ple Aldo Moro 2, 00185, Rome, Italy}

\author {Najla Said}
\affiliation{Physics Department and INFN, Universit\`a di Roma ``La Sapienza'', Ple Aldo Moro 2, 00185, Rome, Italy}

\begin{abstract}
The Atacama Cosmology Telescope (ACT) and the South Pole Telescope (SPT)  have recently provided new, very precise measurements of the cosmic microwave background (CMB) anisotropy damping tail. The values of the cosmological parameters inferred from these measurements, 
while broadly consistent with the expectations of the standard cosmological model,
are providing interesting possible indications for new physics that are definitely worth of investigation.
The ACT results, while compatible with the standard expectation of three neutrino families, indicate a level
of CMB lensing, parametrized by the lensing amplitude parameter $A_{L}$, that is about $70 \%$ higher than expected. 
If not a systematic, an anomalous lensing amplitude could be produced by modifications
of general relativity or coupled dark energy.  
Vice-versa, the SPT experiment, while compatible with a standard level of CMB lensing, prefers
 an excess of dark radiation, parametrized by the effective number of relativistic degrees of freedom
 $N_\mathrm{eff}$. Here we perform a new analysis of these experiments allowing simultaneous
 variations in both these non-standard parameters. We also combine these experiments,
 for the first time in the literature, with the recent WMAP9 data, one at a time.
Including the Hubble Space Telescope (HST) prior on the Hubble constant and information from baryon acoustic oscillations (BAO) surveys provides the following constraints from ACT: $N_\mathrm{eff}=3.23\pm0.47$, $A_{L}=1.65\pm0.33$ at $68 \%$ c.l., while for SPT we
 have $N_\mathrm{eff}=3.76\pm0.34$, $A_{L}=0.81\pm0.12$ at $68 \%$ c.l.. 
In particular, the $A_{L}$ estimates from the two experiments, even when a variation in $N_\mathrm{eff}$ is
allowed, are in tension at more than $95 \%$ c.l..
\end{abstract}


\maketitle

\section{Introduction} \label {sec:intro}

The new measurements of the Cosmic Microwave Background (CMB) anisotropies provided by the
Atacama Cosmology Telescope (ACT) \cite{act2013} and the 
South Pole Telescope (SPT) \cite{spt2013} have both provided
new and exquisitely precise observations of the CMB damping tail.

This angular region of the CMB angular spectra, corresponding to the multipole range going from $\ell \sim 700$ up
to $\ell \sim 3000$, plays a key role in the determination of crucial parameters
like the relativistic number of degrees of freedom $N_\mathrm{eff}$, the
primordial Helium abundance $Y_p$ and the running $dn/d\ln k$ of the scalar spectral index.

Among those parameters, $Y_p$ can be determined unambigously assuming standard
Big Bang Nucleosynthesis (and thus does not represent a free parameter of the theory),  while $dn/d\ln k$ is expected to be negligible in most
inflationary models. On the other hand, the effective number of relativistic degrees of freedom
$N_\mathrm{eff}$ practically parametrizes the energy density of relativistic particles in the early Universe.
In the standard scenario, with the three active relativistic neutrino species,
a value of $N_\mathrm{eff}=3.046$ is expected \cite{Mangano:2005cc}. Deviations from this value due to a non-vanishing neutrino chemical potential are possible but
bound to be small, especially in light of the recent evidences for a large value of the neutrino mixing angle $\theta_{13}$,  see e.g. \cite{Mangano:2011ip,Castorina:2012md}. Thus a detection of $N_\mathrm{eff} \neq 3.046$ would point to 
the presence of physics beyond the standard model of particle physics, like the existence of a yet unknown particle, e.g., a sterile neutrino.

The damping tail is not only affected by those parameters but also
by other physical effects generally taking place at a much later epoch,
well after recombination. These includes, for example, the extragalactic foreground emission of
point sources, radio galaxies, the Sunyaev-Zel'dovich effect and similar unresolved backgrounds.
These foregrounds can however be well identified by their spectral and
angular dependence and have in general a minimal correlation with the 
cosmological parameters.

More importantly, the CMB damping tail is affected by the lensing of CMB photons by dark matter clumps
along the line of sight. This effect is linear, can be computed precisely and 
depends on the same cosmological parameters that affect the primary CMB
spectrum. However, the lensing amplitude is strictly dependent from
the growth of perturbations. This quantity can be significant different
if, for example, general relativity is not the correct theory to describe
gravity at the very large scales. If the accelerated expansion of
our universe is indeed provided not by a dark energy component but by  
modified gravity, the perturbation growth could be dramatically
different and change the expectations of lensing (see, for example
\cite{calabrese1} and references therein). In order to test the
correct amplitude of the lensing signal, one can introduce a calibration
parameter $A_L$, as in \cite{calabrese2},  that scales the lensing potential in such a way that 
$A_L=0$ corresponds to the complete absence of lensing, while $A_L=1$ is the expected lensed result assuming
general relativity. A robust detection of $A_L$ being different from unity would hint to the fact that 
general relativity is not the correct theory to describe gravity at the cosmological scales.

The new ACT and SPT data, while broadly consistent with the expectations
of the standard $\Lambda$CDM scenario, are indeed providing interesting hints
for deviations from the simplest $\Lambda$CDM model when combined with the results from 7 years of observations
from the WMAP satellite (WMAP7, \cite{wmap7}).

The SPT experiment, for example, is confirming an indication for a value
for $N_\mathrm{eff}>3.046$. This indication, already present in the previous
data release (see e.g. \cite{hou}, \cite{archidiacono} and \cite{giusarma}), is
marginal when considering only the WMAP7+SPT data with
$N_\mathrm{eff}=3.62\pm0.48$ at $68 \%$ c.l.. However, it is more  
significant when the SPT data is combined with the measurement of the Hubble constant
$H_0=73.8\pm2.4 \mathrm{km}\,\mathrm{s}^{-1}\,\mathrm{Mpc}^{-1}$ from the Hubble Space Telescope (HST) \cite{hst} and with information from 
Baryonic Acoustic Oscillation (BAO) data (see Table 4 in \cite{spt2013}), yielding
a final value of $N_\mathrm{eff}=3.71\pm0.35$.

At the same time, the ACT collaboration presented a similar analysis obtaining
different results. In particular, the WMAP7+ACT data alone constrain the neutrino
number to be $N_\mathrm{eff}=2.78\pm0.55$, i.e. perfectly consistent with the standard
three-neutrino framework. When the ACT data is combined with HST and BAO data the
value is higher, $N_\mathrm{eff}=3.52\pm0.39$, but still consistent with 
three neutrinos families (see Table III in \cite{act2013}).

Interestingly, this is not the only tension between the two datasets.
If we now consider the results on the lensing amplitude parameter, the SPT dataset
is fully compatible with the standard expectation, with $A_L=0.86^{+0.15}_{-0.13}$
at $68 \%$ c.l. (see \cite{sptexp}), while the ACT data suggest a 2$\sigma$ deviation
from the standard expectation, with $A_L=1.70 \pm 0.38$ at $68 \%$ c.l..

In this brief paper we further investigate these discrepancies by improving
these analyses in two ways. First of all, we perform our analyses allowing
{\it both} $N_\mathrm{eff}$ and $A_L$ parameters to vary at the same time.
As we will see, this let to better identify the tension between the two
experiments. Secondly, we add the recent dataset from nine years of observations
coming from the WMAP satellite as in \cite{wmap9}. Both ACT and
SPT teams used the previous 7-year WMAP dataset in their papers and some,
albeit small, differences are present when the updated dataset 
is considered.

Our paper is simply organized as follows: in the next section we describe
the analysis method, in Section III we present our results and
in Section IV we derive our conclusions.

\section{Data Analysis Method} \label{sec:method}

Our analysis is based on a modified version of the public
CosmoMC \cite{Lewis:2002ah} Monte Carlo Markov Chain code. We consider
the following CMB data: WMAP9 \cite{wmap9}, SPT \cite{spt2013}, 
ACT \cite{act2013} including measurements up to a 
maximum multipole number of $l_{\rm max}=3750$.
For all these experiments we make use of the publicly available
codes and data. For the ACT experiment we use the "lite" 
version of the likelihood \cite{dunkleyact}. Since ACT and SPT dataset are providing different results on the
parameters, we will consider them separately. Thus our basic CMB-only datasets consist of the WMAP9+ACT and WMAP9+SPT data.

We also consider the effect of including additional dataset to the basic datasets just described.
Consistently with the measurements of the HST \cite{hst}, we consider a gaussian prior on the Hubble constant 
$H_0=73.8\pm2.4 \,\mathrm{km}\,\mathrm{s}^{-1}\,\mathrm{Mpc}^{-1}$. We also include information from 
measurements of baryonic acoustic oscillations (BAO) from galaxy surveys. 
Here we follow the approach presented in \cite{wmap9} combining
four datasets: 6dFGRS from \cite{beutler/etal:2011},
SDSS-DR7 from \cite{padmanabhan/etal:2012}, SDSS-DR9
from \cite{anderson/etal:2012} and WiggleZ from \cite{blake/etal:2012}.

We sample the standard six-dimensional set of cosmological parameters,
adopting flat priors on them: the baryon and cold dark matter densities
$\Omega_{\rm b} h^2$ and $\Omega_{\rm c} h^2$, the ratio of the sound horizon 
to the angular diameter distance at decoupling $\theta$, the optical 
depth to reionization $\tau$, the scalar spectral index $n_s$, the 
overall normalization of the spectrum $A_s$ at $k=0.002\Mpc^{-1}$.
 
Since the ACT and SPT data are showing indications for deviations from their
standard values, we also consider variations in the effective number of relativistic 
degrees of freedom $N_\mathrm{eff}$ and in the lensing amplitude parameter $A_L$
as defined in \cite{calabrese2}, that simply rescales the lensing potential:
\begin{equation}
C_{\ell}^{\phi \phi} \rightarrow A_L C_{\ell}^{\phi \phi}
\end{equation}

\noindent where $C_{\ell}^{\phi \phi}$ is the power spectrum of the
lensing field. We take flat priors on all the parameters; in particular, we take $1 < N_\mathrm{eff} < 10$ and $0 < A_{L} < 4$. 

In our basic runs, we do not consider the effect of massive neutrinos. We perform additional runs in which
we allow for a non-vanishing neutrino mass, parametrized by means of the neutrino fraction $f_\nu \equiv \Omega_\nu / \Omega_c$.
We always assume standard Big Bang Nucleosynthesis, so that the  Helium abundance $Y_p$ is uniquely determined
by the values of $\Omega_b h^2$ and $N_\mathrm{eff}$.

Finally, in order to assess the convergence of our MCMC chains, we compute the Gelman and Rubin $R-1$ parameter
demanding that $R-1 < 0.03$.

\section{Results}\label{sec:results}

\begin{figure*}[htb!]
\includegraphics[width=\columnwidth]{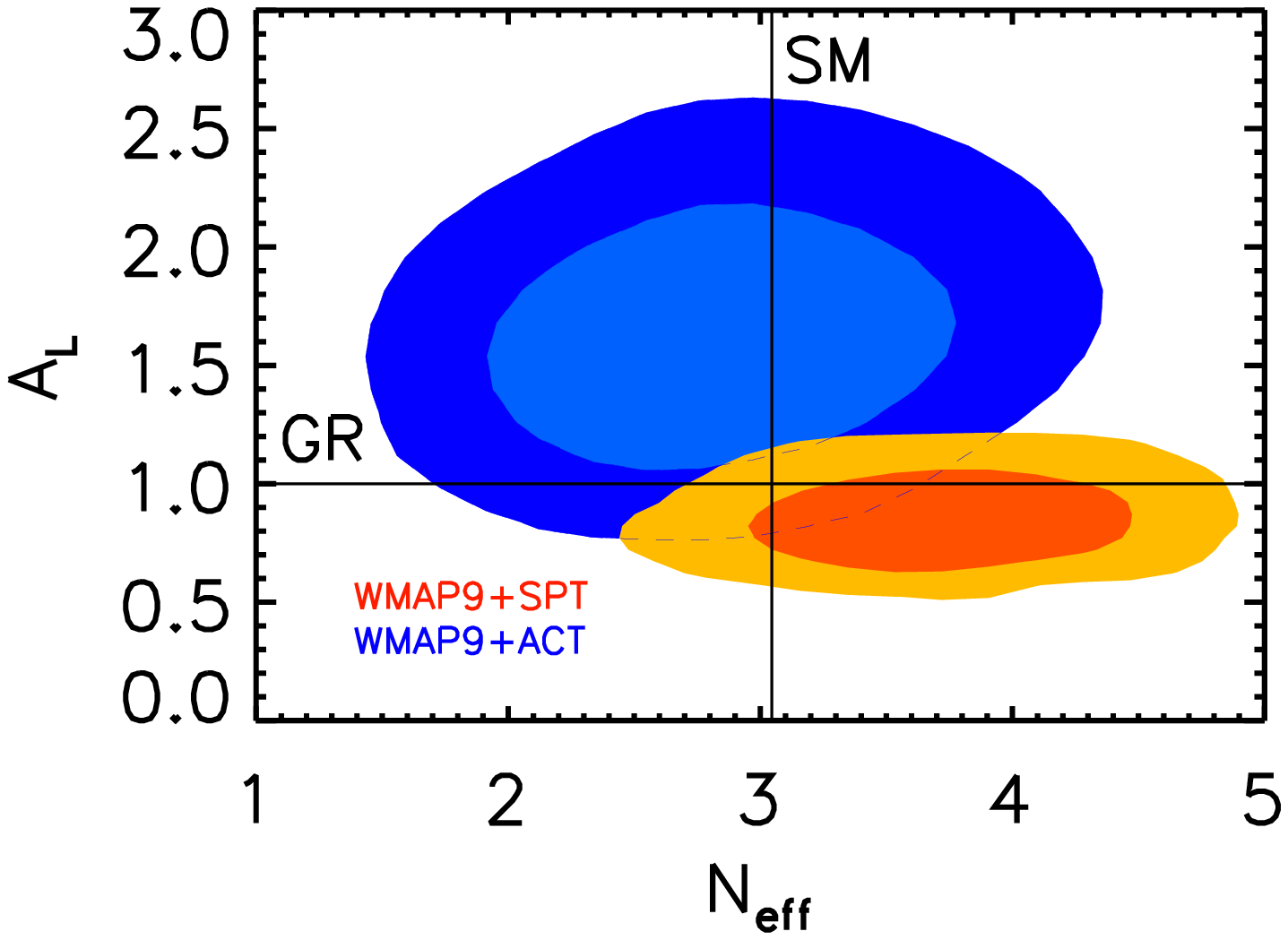}
\includegraphics[width=\columnwidth]{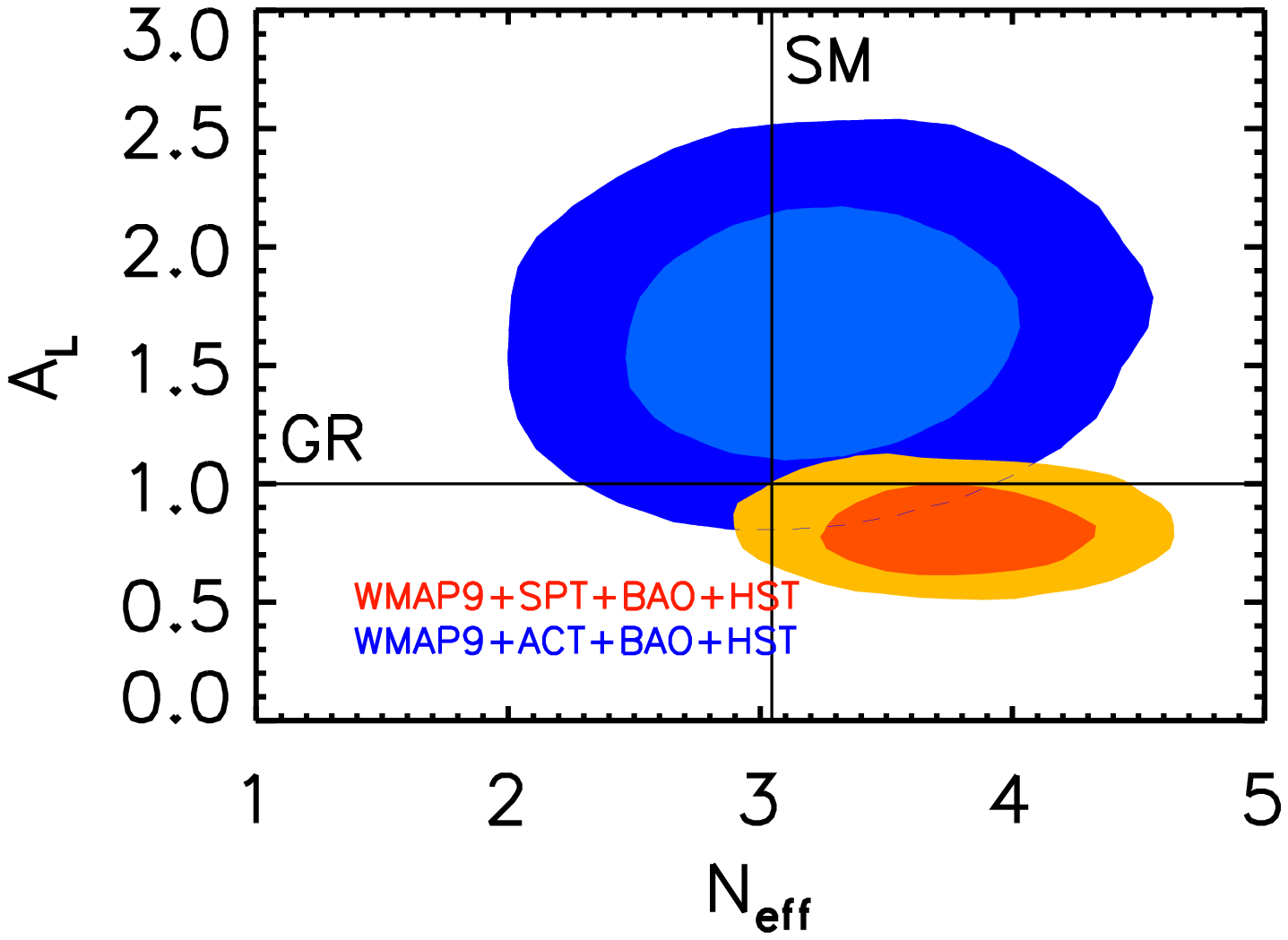}
\caption{Constraints in the $A_L$ - $N_\mathrm{eff}$ plane 
from a CMB only analysis (left panel) and including the HST prior
and BAO (right panel). The blue contour includes the ACT data
while the red contour refers to the SPT data. The line at $A_L=1$ indicates the standard
expectations based on General Relativity. The line at $N_\mathrm{eff}=3.046$
indicates the prediction from the standard model with three neutrino flavours.}  
\label{alneff}
\end{figure*}

\begin{figure*}[htb!]
\includegraphics[width=\columnwidth]{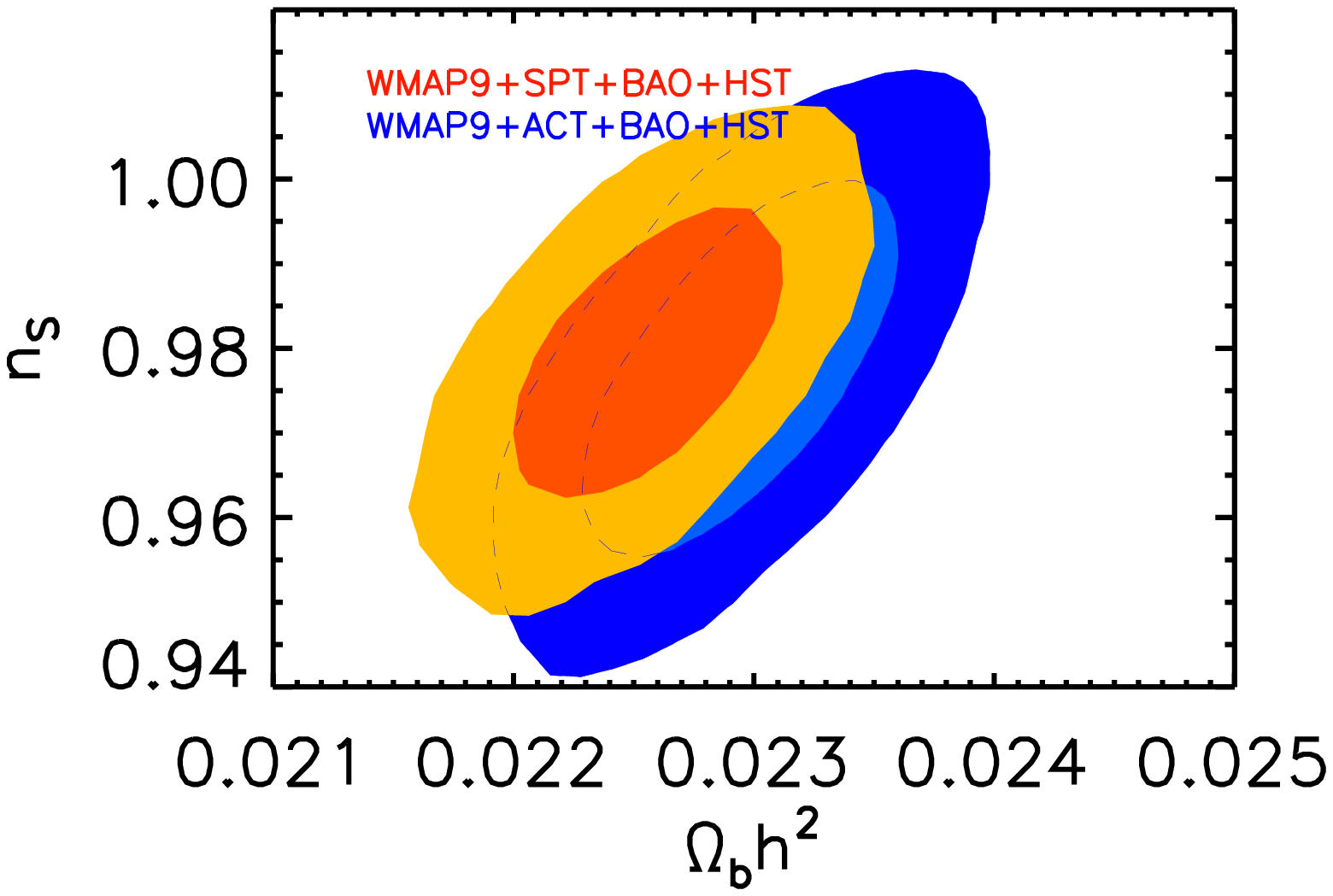}
\includegraphics[width=\columnwidth]{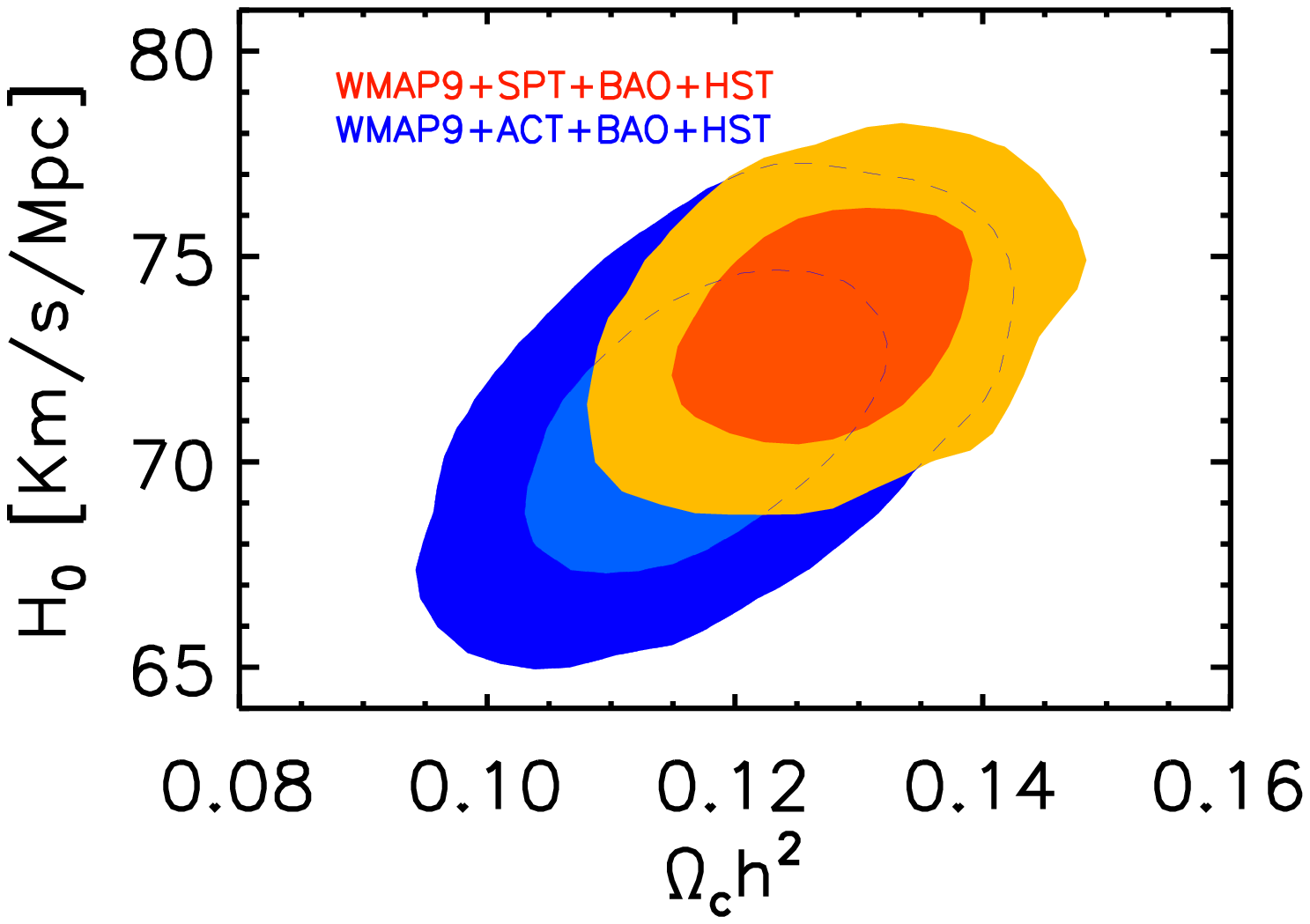}
\caption{Constraints in the $\Omega_b h^2$ - $n_s$ plane 
(left panel) and on the $\Omega_ch^2$-$H_0$ plane (right panel)
from ACT (blue contours) and SPT (orange contours) including WMAP9, HST
and BAO data. The ACT-SPT tension is less pronounced for these parameters.}  
\label{altri}
\end{figure*}

\begin{table*}[htb!]
\begin{center}
\begin{tabular}{|l||c||c|c||c|}
\hline
\hline
 Parameters & {\bf SPT}+WMAP9 & {\bf ACT}+WMAP9 & {\bf SPT}+WMAP9+HST+BAO & {\bf ACT}+WMAP9+HST+BAO\\
\hline
\hline
$\Omega_b h^2$ & $0.02264\pm0.00051$ & $0.02283\pm0.00052$ & $0.02255\pm0.00036$ & $0.02294\pm0.00042$\\
$\Omega_c h^2$ & $0.1232\pm0.0080$ & $0.110\pm0.010$ & $0.1274\pm0.0075$ & $0.1178\pm0.0094$\\
$\theta$ & $1.0415\pm0.0012$ & $1.0412\pm0.0025$ & $1.0413\pm0.0012$ & $1.0400\pm0.0024$\\
$\tau$ & $0.088\pm0.014$ & $0.090\pm0.014$ & $0.085\pm0.013$ & $0.090\pm0.014$\\
$n_s$ & $0.982\pm0.018$ & $0.969\pm0.019$ & $0.979\pm0.011$ & $0.978\pm0.014$\\
$N_\mathrm{eff}$ & $3.72\pm0.46$ & $2.85\pm0.56$ & $3.76\pm0.34$ & $3.23\pm0.47$\\
$A_L$ & $0.85\pm0.13$ & $1.64\pm0.36$ & $0.81\pm0.12$ & $1.65\pm0.33$\\
$H_0 [\mathrm{km}/\mathrm{s}/\mathrm{Mpc}]$ & $74.6\pm3.7$ & $69.9\pm3.7$ & $73.3\pm1.8$ & $71.1\pm2.4$\\
$\log(10^{10} A_s)$ & $3.169\pm0.048$ & $3.174\pm0.045$ & $3.185\pm0.034$ & $3.174\pm0.037$\\
$\Omega_{\rm \Lambda}$ & $0.736\pm0.023$ & $0.728\pm0.025$ & $0.721\pm 0.015$ & $0.721\pm0.017$\\
$\Omega_{\rm m}$ & $0.264\pm0.023$ & $0.272\pm0.025$ & $0.279\pm0.015$ & $0.279\pm0.017$\\
Age/Gyr & $13.14\pm0.43$ & $13.90\pm0.55$ & $13.12\pm0.28$ & $13.55\pm0.41$\\
$D_{3000}^{SZ}$ & $5.8\pm2.4$ & ---  & $5.8\pm2.5$ & ---\\
$D_{3000}^{CL}$ & $5.2\pm2.1$ & --- & $5.3\pm2.1$ & ---\\
$D_{3000}^{PS}$ & $19.6\pm2.5$ & --- & $19.5\pm2.5$ & ---\\
$A_{SZ}$ & --- & $1.00\pm0.57$ & --- & $0.91\pm0.57$\\
\hline
$\chi^2_{\rm min}$ & $3806.25$ & $3798.83$ & $3808.06$ & $3800.59$ \\
\hline
\hline
\end{tabular}
\caption{Cosmological parameter values and $68 \%$ confidence level errors. The SPT and ACT datasets produce
different values for some of the parameters, most notably $N_\mathrm{eff}$ and $A_L$.}
\label{standard}
\end{center}
\end{table*}

\begin{table*}[htb!]
\begin{center}
\begin{tabular}{|l||c||c|c||c|}
\hline
\hline
 Parameters & {\bf SPT} & {\bf SPT} & {\bf ACT}& {\bf ACT}\\
  & +WMAP9+HST+BAO & +WMAP9+HST+BAO & +WMAP9+HST+BAO& +WMAP9+HST+BAO\\
\hline
\hline
$\Omega_b h^2$ & $0.02278\pm0.00036$ & $0.02274\pm0.00037$ & $0.02289\pm0.00042$ & $0.02303\pm0.00043$\\
$\Omega_c h^2$ & $0.1305\pm0.0084$ & $0.1310\pm0.0082$ & $0.1163\pm0.0087$ & $0.1193\pm0.0097$\\
$\theta$ & $1.0412\pm0.0011$ & $1.0412\pm0.0011$ & $1.0402\pm0.0023$ & $1.0402\pm0.0023$\\
$\tau$ & $0.089\pm0.013$ & $0.089\pm0.013$ & $0.094\pm0.014$ & $0.091\pm0.014$\\
$n_s$ & $0.989\pm0.012$ & $0.987\pm0.012$ & $0.976\pm0.014$ & $0.981\pm0.014$\\
$N_\mathrm{eff}$ & $3.89\pm0.38$ & $3.88\pm0.37$ & $3.09\pm0.43$ & $3.28\pm0.48$\\
$\Sigma m_{\nu} [eV]$ & $0.40\pm0.22$ & $< 0.77$ ($95 \%$ c.l.) & $<0.40$ ($95 \%$ c.l.)& $<0.53$ ($95 \%$ c.l.)\\
$A_L$ & $1.00$ & $0.90\pm0.15$ & $1.00$ & $1.78\pm0.38$\\
$H_0 [\mathrm{km}/\mathrm{s}/\mathrm{Mpc}]$ & $72.4\pm2.0$ & $72.5\pm1.9$ & $69.3\pm2.4$ & $70.1\pm2.5$\\
$\log(10^{10} A_s)$ & $3.156\pm0.034$ & $3.166\pm0.036$ & $3.177\pm0.036$ & $3.162\pm0.037$\\
$\Omega_{\rm \Lambda}$ & $0.707\pm0.020$ & $0.707\pm0.019$ & $0.710\pm 0.019$ & $0.710\pm0.020$\\
$\Omega_{\rm m}$ & $0.293\pm0.020$ & $0.293\pm0.019$ & $0.290\pm0.019$ & $0.290\pm0.020$\\
Age/Gyr & $13.12\pm0.29$ & $13.10\pm0.28$ & $13.75\pm0.40$ & $13.59\pm0.42$\\
$D_{3000}^{SZ}$ & $5.8\pm2.4$ & $6.1\pm2.4$ & ---  & ---\\
$D_{3000}^{CL}$ & $5.3\pm2.2$ & $5.2\pm2.1$& --- & ---\\
$D_{3000}^{PS}$ & $19.3\pm2.4$ &$19.3\pm2.5$ & ---  & ---\\
$A_{SZ}$ & --- & ---& $0.97\pm0.57$ & $0.97\pm0.57$\\
\hline
$\chi^2_{\rm min}/2$ & $3808.0$ & $3807.5$& $3802.47$ & $3800.59$ \\
\hline
\hline
\end{tabular}
\caption{Cosmological parameter values and $68 \%$ confidence level errors for the analysis that
considers massive neutrinos. As we can see, varying $A_L$ strongly affects the constraints on the
total neutrino mass. Vice-versa, allowing for a neutrino mass renders the SPT value for $A_L$ more 
compatible with the standard value while exacerbates the problem for the ACT dataset.}
\label{standard}
\end{center}
\end{table*}

\begin{figure*}[htb!]
\includegraphics[width=\columnwidth]{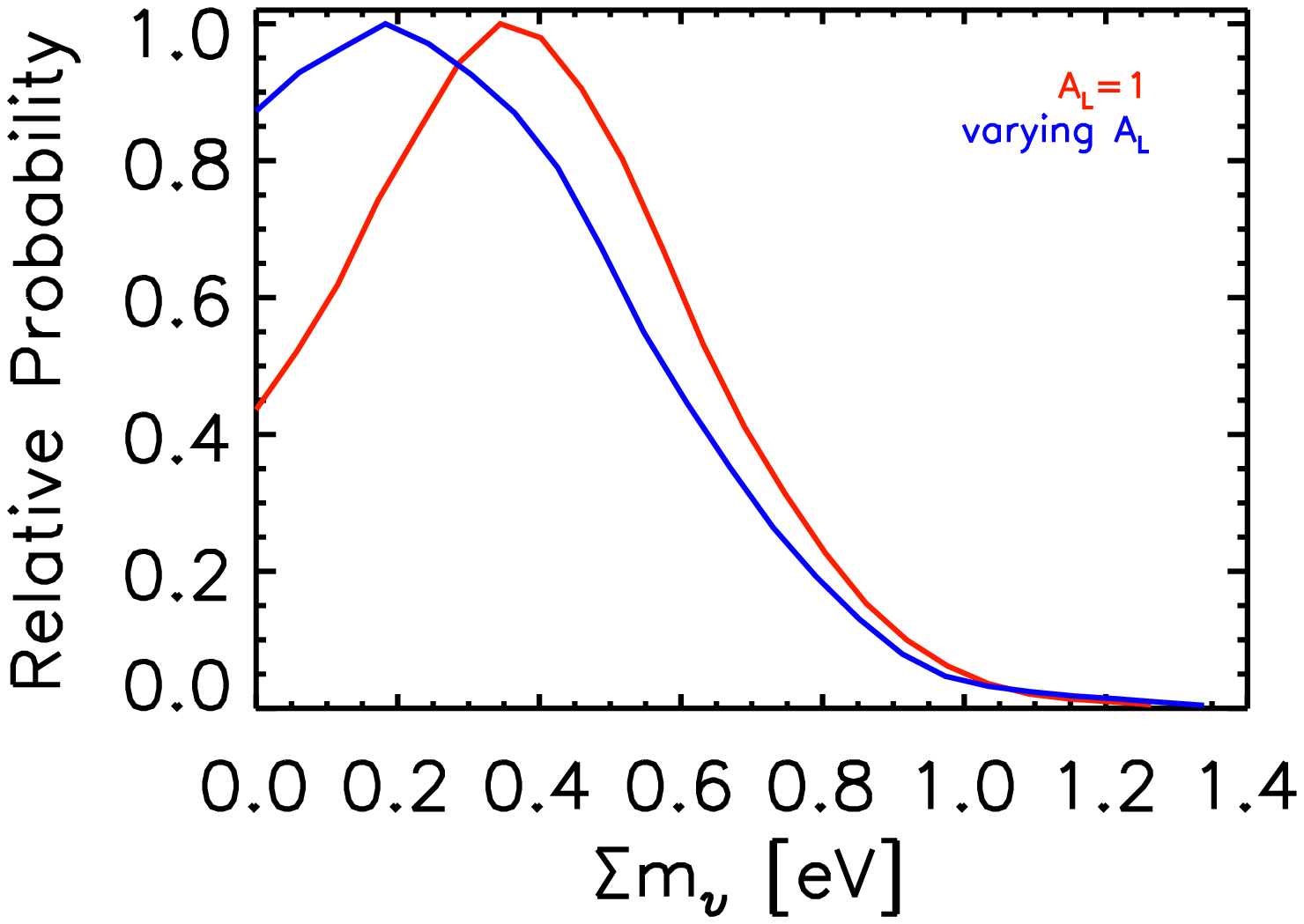}
\includegraphics[width=\columnwidth]{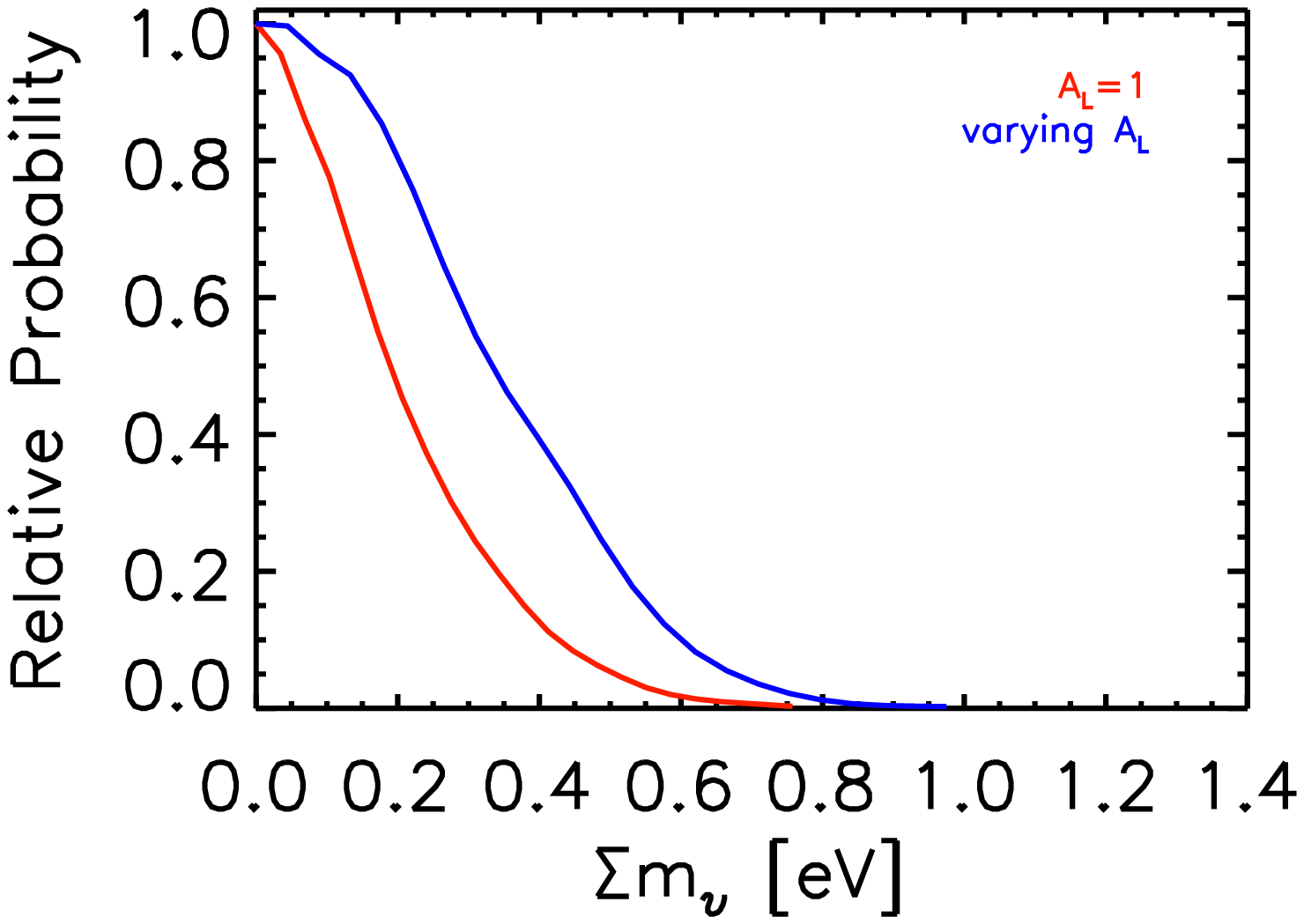}
\caption{Posterior distribution function for the total neutrino mass
parameter $\Sigma m_{\nu}$ from a SPT+WMAP+BAO+HST analysis  
(left panel) and ACT+WMAP+BAO+HST (right panel)
in the case of fixing lensing to $A_L=1$ and letting it to vary.
As we can see, if we let the $A_L$ parameter to vary the small indication
for a neutrino mass from the SPT analysis vanishes. At the same time, letting the $A_L$ parameter to
vary weakens the constraints from ACT.}  
\label{posteriormnu}
\end{figure*}

\begin{figure}[ht]
\includegraphics[width=\columnwidth]{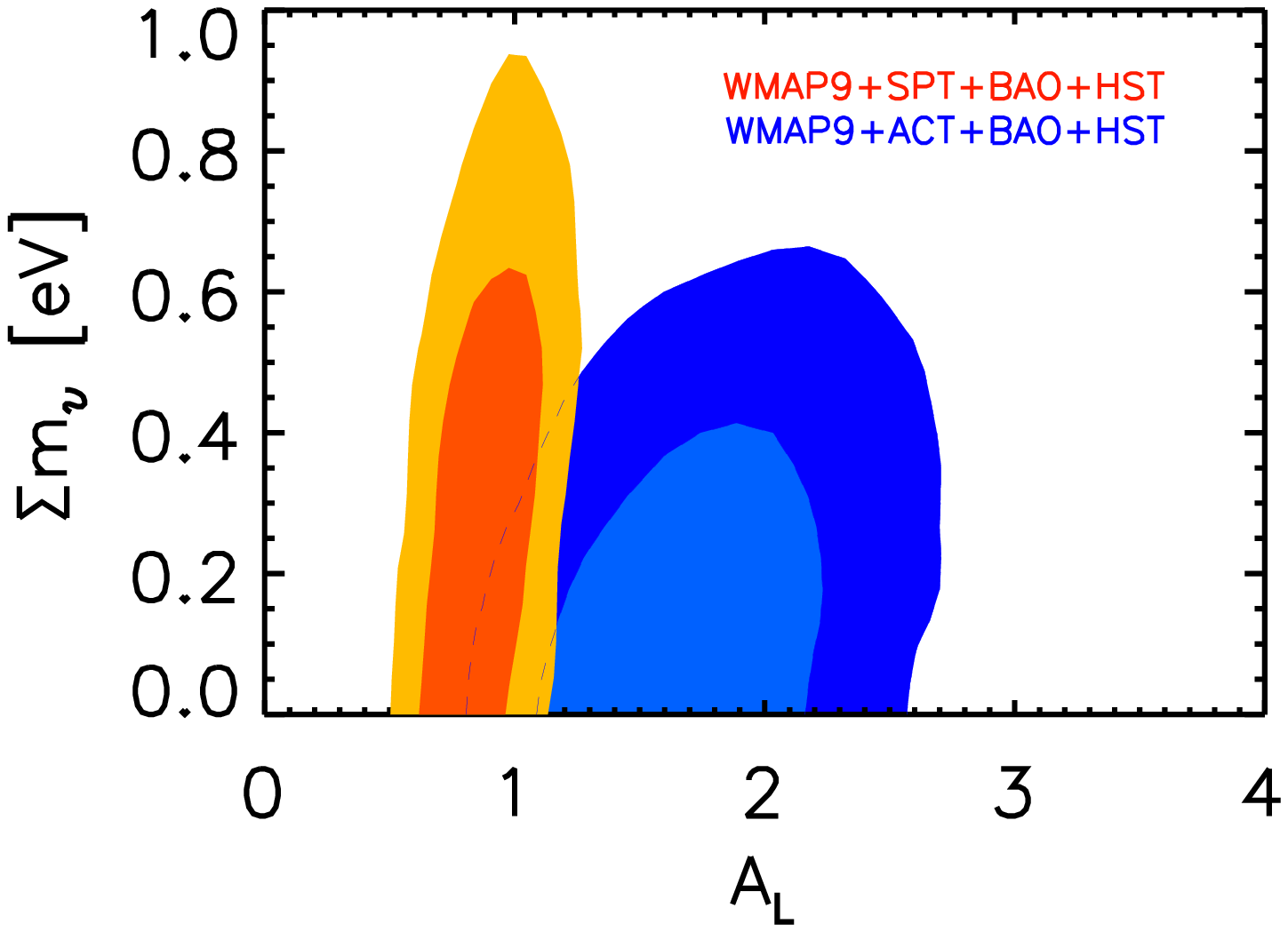}
\caption{Constraints in the $A_L$ vs $\Sigma m_{\nu}$ plane for the 
SPT+WMAP+BAO+HST and ACT+WMAP+BAO+HST datasets. A degeneracy is present between the
two parameters: larger values for $A_L$ let larger neutrino masses to be more consistent
with the data. The SPT indication for a neutrino mass is driven by the low value of $A_L$ obtained
in the neutrino massless case.}  
\label{almnu}
\end{figure}

As stated in the previous section, we consider the ACT and SPT datasets separately.
We therefore perform the following four analyses: {WMAP9$+$ACT}, {WMAP9$+$ACT$+$HST$+$BAO}, 
{WMAP9$+$SPT}, and {WMAP9$+$SPT$+$HST$+$BAO}.

In Table I we report the constraints on the considered parameters from each run.
As we can see, the ACT and SPT are providing  significantly different constraints on the $N_\mathrm{eff}$ and $A_L$ parameters.

In order to further investigate this discrepancy, we plot in figure \ref{alneff} 
the 2-D constraints on the $N_\mathrm{eff}$ vs $A_L$ plane for the
CMB only case and for the CMB+HST+BAO analysis.

As we can see, a tension is clearly present since 
the central values for $N_\mathrm{eff}$ and $A_L$ obtained from WMAP9+ACT analysis are outside
the $95 \%$ confidence level of the WMAP9+SPT and vice-versa.
Namely, the ACT dataset is pointing towards a value of $N_\mathrm{eff}$ fully consistent
with the standard scenario of $N_\mathrm{eff}=3.046$, while (as it can be seen from Table 1 and Figure 1), 
preferring at the same time an exotic high value for the lensing potential, with $A_L$ larger than
 unity at more than $95\%$ c.l. when the BAO and HST datasets are included. 
Considering the $95 \%$ confidence levels we found $A_L=1.64_{-0.56}^{+0.63}$
for the WMAP+ACT analysis and $A_L=1.65_{-0.52}^{+0.56}$ for the WMAP+ACT+BAO+HST.

The situation is opposite for the SPT data: while SPT is fully consistent with 
$A_L=1$, $N_\mathrm{eff}$ is constrained to a larger value than the standard
expectation. When also the HST and BAO data are included, we see that not only
a value of $N_\mathrm{eff}>3.04$ is suggested at more than $95 \%$ c.l., but also
a value of $A_L$ smaller than one is suggested at about $68 \%$ c.l..

In particular, we found that $A_L <1.07$ at $95 \%$ c.l. from WMAP9+SPT+BAO+HST
while $A_L> 1.13$ at $95 \%$ c.l. from WMAP9+ACT+BAO+HST, i.e. for the
lensing parameter the SPT and ACT datasets are providing constraints that are
in disagreement at more than $95 \%$ c.l..

It is interesting to note that the tension between the ACT and SPT datasets is
clearly not limited to $A_L$ or $N_\mathrm{eff}$: also the constraints on $H_0$,
$n_s$, $\Omega_b h^2$ and $\Omega_c h^2$ appear as quite different. The discrepancy is however less
significant since the central values are inside the $95 \%$ confidence level
of each analysis (see Figure \ref{altri}).

The results discussed so far are relative to the analysis in which all neutrinos are considered as relativistic and massless.
Since the SPT dataset is claiming a detection at $95 \%$ c.l. for a neutrino mass with 
$\Sigma m_{\nu}=0.48 \pm 0.21$ in a WMAP7+SPT+BAO+HST analysis (see \cite{spt2013}), 
it is clearly interesting to consider also massive neutrinos.

In table II we present the constraint on cosmological parameters from 
the WMAP9+SPT+HST+BAO and WMAP9+ACT+HST+BAO datasets 
respectively when variation in the neutrino masses are included in
two cases: varying $A_L$ and fixing $A_L=1$.

As we can see, while the ACT dataset does not favour the presence of
neutrino masses, the SPT dataset gives $\Sigma m_{\nu}=0.40 \pm 0.22$
at $68 \%$ c.l. in the case of $A_L=1$ and a lower limit limit $\Sigma m_{\nu}> 0.04 eV$
at $95 \%$ c.l.. This is consistent with the
results reported in \cite{spt2013} considering the different WMAP
and BAO datasets. However, when $A_L$ is let to vary, the
evidence for a neutrino mass vanishes, as also clearly see in Figure
\ref{posteriormnu}.

We can better see what is happening by looking at the constraints
in the $A_L$ vs $\Sigma m_{\nu}$ plane in Figure 4.
As we can see, there is a degeneracy between $A_L$ and
$\Sigma m_{\nu}$. Namely, a larger value of $\Sigma m_{\nu}$ decreases the lensing 
signal and can be compensated with a larger $A_L$. Since the SPT dataset is preferring smaller
values of the lensing parameter, an analysis with $A_L=1$  forces
the neutrino mass to be more consistent with the data.

Is also worth mentioning that including a neutrino mass exacerbates the 
lensing problem for ACT. The lensing parameter $A_L$ is even higher
when massive neutrinos are considered (see Table II).

\section{Discussion}

In this paper we have pointed out a tension between the parameter values 
estimated from the recent ACT and SPT datasets.
This discrepancy, albeit not significantly more than the $95 \%$ confidence level, is indicating
the possible presence of systematics in at least one of the two datasets.
The SPT experiment is confirming the previous indications for a "dark radiation" 
component with $N_\mathrm{eff}=3.76\pm0.34$ at $68 \%$ c.l.; in particular we have found that
$N_\mathrm{eff}> 3.08$ at more than  $95\%$ c.l..
This result is clearly interesting since, if confirmed with larger significance
by future data, could be possibly explained by several physical mechanisms, and would
hint to new physics. In fact, the most conventional explanation for $N_\mathrm{eff}>3.046$ would be the
presence of non-vanishing neutrino chemical potentials, i.e. of a cosmological lepton asymmetry.
However, as it was shown in Refs.\cite{Mangano:2011ip,Castorina:2012md} through the analysis of BBN and CMB data, lepton asymmetries can at most account for $N_\mathrm{eff}\simeq 3.1$, given
the recent measurements of the neutrino mixing angle $\theta_{13}$ by the Daya Bay  \cite{An:2012eh} and RENO experiments \cite{Ahn:2012nd} 
that exclude a zero value for $\theta_{13}$ with high significance.

Thus, if confirmed, a value of $N_\mathrm{eff}$ larger than 3.1 definitely requires some non-conventional explanation. 
Sterile neutrinos, extra dimensions, gravity waves or non-standard neutrino
decoupling could all be viable new mechanisms to explain a value of $N_\mathrm{eff}$
larger than the standard value (see e.g. \cite{theories}).

The ACT experiment is, on the contrary, fully consistent with $N_\mathrm{eff}=3.04$
even when the HST and BAO dataset are included. There is clearly no evidence
for dark radiation from ACT; moreover the case for a fourth, 
massless neutrino, such that $N_\mathrm{eff}\simeq 4$ is excluded at more than $95 \%$ c.l.
when only CMB data is considered. In particular, we found at $95 \%$ c.l. that 
$N_\mathrm{eff}=2.85_{-0.91}^{+0.95}$ for WMAP9+ACT and $N_\mathrm{eff}=3.23_{-0.76}^{+0.77}$
for WMAP9+ACT+BAO+HST.
It is interesting to notice that our WMAP9+ACT+BAO+HST run provides the
constraint $N_\mathrm{eff}=3.23\pm0.47$ while a similar analysis from ACT gives 
$N_\mathrm{eff}=3.52\pm0.39$ but with $A_L=1$ and with the WMAP7 data.

However, ACT presents a value for the lensing parameter that is off by
more than $95 \%$ from the expected value $A_L=1$. This result is 
probably more difficult to explain from a physical point of view than a 
deviation in $N_\mathrm{eff}$ and calls for more drastic changes in the 
cosmological model. A possible way to enhance the lensing signal is 
to assume a modification to general relativity. $f(R)$ models as those 
investigated in \cite{calabrese1} could in principle enhance the lensing
signal, even if it is not clear if they could enhance it by $\sim 70 \%$
and be at the same time consistent with other independent limits coming 
from tests of general relativity, like, e.g. solar system tests.
Other possible explanations include coupled dark energy models (see e.g. \cite{coupled}
and references therein).
Clearly, it may be that the ACT lensing signal is on the contrary simply produced by some 
unknown systematic as also suggested by the inclusion of the 
ACT deflection spectrum data, that shifts the value to 
$A_L=1.3\pm0.23$ (\cite{act2013}). However it is not clear if
this systematic could also affect the ACT constrain on $N_\mathrm{eff}$
and other parameters.

The SPT experiment is compatible with $A_L=1$ but is suggesting a
value $A_L<1$ at about $68 \%$ c.l. especially when also the BAO and HST
data are included. 

The ACT and SPT measurements of $A_L$, even if we consider variation in
the $N_\mathrm{eff}$ parameter, are in disagreement at more than $95 \%$ c.l..

Finally, we have also considered variation in the neutrino mass and show
that the current indication for a neutrino mass from the SPT+WMAP9+BAO+HST run
is driven by the lower lensing amplitude measured by SPT. If we let the lensing
parameter $A_L$ to vary the indication for a neutrino
mass vanishes. 
Moreover, we have shown that the inclusion of a neutrino mass exacerbates the
lensing problem for the ACT data with the $A_L$ even more discrepant with
the $A_L=1$ case. The constraints on the neutrino mass from ACT are weaker when
variations in $A_L$ are considered. 

In this paper we have only considered a limited set of parameters but the tension
between SPT and ACT is present also in other, relevant, parameters.
The SPT dataset, for example, shows a preference for a negative running of the
inflationary spectral index at more than $95 \%$ c.l. while the ACT data 
is consistent with a zero running in between the $95 \%$ c.l. (see Figure 11
of \cite{act2013}).

We therefore conclude that the whole picture is, at the moment, 
stimulating and puzzling at the same time. The ACT and SPT collaborations
have provided an impressive confirmation of the theoretical
expectations concerning the damping tail of the CMB anisotropy spectrum. However they are 
also suggesting interesting deviations from the standard picture, that are unfortunately very 
different and opposite.
It will be the duty of future re-analyses of the ACT and SPT data
(possibly stemming from within the collaborations themselves) and experiments (e.g., Planck)
to finally decide whether what ACT and SPT are
currently seeing is due to dark radiation, dark gravity or
more simply to an unidentified (hence, dark too) experimental systematic
effect.

\
\

\section{Acknowledgements}

It is a pleasure to thank Andrea Marchini and Valentina Salvatelli for help. 
We also acknowledge CASPUR for computational support. The work of ML has been supported by Ministero dell'Istruzione, dell'Universit\`a e 
della Ricerca (MIUR) through the PRIN grant  ``Galactic and extragalactic polarized microwave emission'' (contract number PRIN 2009XZ54H2-002).
\
\
\

\end{document}